\documentclass[twocolumn,floatfix,superscriptaddress,pra,aps,10pt,longbibliography]{revtex4-2}
\usepackage{times}
\usepackage{amssymb}
\usepackage{color}
\usepackage{amsmath}
\usepackage{mathtools}
\usepackage{amsbsy}
\usepackage{amsthm}
\usepackage{graphicx}
\usepackage{bbm}
\usepackage{bm}
\usepackage{epsfig}
\usepackage{xfrac}
\usepackage{xcolor}
\usepackage{enumerate}
\usepackage{multirow}
\usepackage{soul}
\usepackage[normalem]{ulem}
\definecolor{myurlcolor}{rgb}{0,0,0.7}
\definecolor{myrefcolor}{rgb}{0.8,0,0}
\usepackage[unicode=true,pdfusetitle, bookmarks=true,bookmarksnumbered=false,bookmarksopen=false, breaklinks=false,pdfborder={0 0 0},backref=false,colorlinks=true, linkcolor=myrefcolor,citecolor=myurlcolor,urlcolor=myurlcolor]{hyperref}

\newcommand{\braket}[2]{\langle #1|#2\rangle}

\renewcommand{\ss}{{\rm{ss}}}
\renewcommand{\t}[1]{\textrm{#1}}
\newcommand{\tr}[0]{\mathrm{Tr}}

\usepackage{braket}
\newcommand{\thmref}[1]{\hyperref[#1]{Theorem~\ref{#1}}}
\newcommand{\lemmaref}[1]{\hyperref[#1]{Lemma~\ref{#1}}}
\newcommand{\figref}[1]{\hyperref[#1]{Fig.~\ref{#1}}}
\newcommand{\figaref}[1]{\hyperref[#1]{Fig.~\ref{#1}a}}
\newcommand{\figbref}[1]{\hyperref[#1]{Fig.~\ref{#1}b}}
\newcommand{\figcref}[1]{\hyperref[#1]{Fig.~\ref{#1}c}}
\renewcommand{\eqref}[1]{\hyperref[#1]{Eq.~(\ref{#1})}}
\newcommand{\eqsref}[2]{\hyperref[#1]{Eqs.~(\ref{#1})-(\ref{#2})}}
\newcommand{\appref}[1]{\hyperref[#1]{Appx.~\ref{#1}}}

\newcommand{\bb}{b}

\usepackage{orcidlink}
\usepackage{dsfont}

\usepackage{tabularray}

\begin{document}

\title{Time correlations from steady-state expectation values}

\author{Wojciech G{\'o}recki\,
\orcidlink{0000-0001-9912-9186}}
\altaffiliation[Current address: ]{\textit{Freie Universit\"at Berlin, Fachbereich Physik and Dahlem Center for Complex Quantum Systems, Arnimallee 14, 14195 Berlin}}
\affiliation{INFN Sez. Pavia, via Bassi 6, I-27100 Pavia, Italy}
\author{Simone Felicetti\,
\orcidlink{0000-0003-0447-5380}}
\affiliation{Institute for Complex Systems, National Research Council (ISC-CNR), Via dei Taurini 19, 00185 Rome, Italy}
\affiliation{Physics Department, Sapienza University, P.le A. Moro 2, 00185 Rome, Italy}
\author{Lorenzo Maccone\,
\orcidlink{0000-0002-6729-5312}}
\affiliation{INFN Sez. Pavia, via Bassi 6, I-27100 Pavia, Italy}
\affiliation{Dip. Fisica, University of Pavia, via Bassi 6, I-27100 Pavia, Italy}
\author{Roberto Di Candia\,
\orcidlink{0000-0001-9087-2125}}
\affiliation{Department of Information and Communications Engineering, Aalto University, Espoo, 02150 Finland}
\affiliation{Institute for Complex Systems, National Research Council (ISC-CNR), Via dei Taurini 19, 00185 Rome, Italy}
\affiliation{Dip. Fisica, University of Pavia, via Bassi 6, I-27100 Pavia, Italy}

\begin{abstract}
Recovering properties of correlation functions is typically challenging. On the one hand, experimentally, it requires measurements with a temporal resolution finer than the system's dynamics. On the other hand, analytical or numerical analysis requires solving the system evolution. Here, we use recent results of quantum metrology with continuous measurements to derive general lower bounds on the relaxation and second-order correlation times that are both easy to calculate and measure. These bounds are based solely on \emph{steady-state} expectation values and their derivatives with respect to a system parameter, and can be readily extended to the autocorrelation of arbitrary observables. 
We validate our method on two examples of critical quantum systems: a critical driven-dissipative resonator, where the bound matches analytical results for the dynamics, and the infinite-range Ising model, where only the steady state is solvable, and thus the bound provides information beyond the reach of existing analytical approaches. Our results can be applied to the experimental characterization of ultrafast systems and to the theoretical analysis of many-body models whose dynamics are hard to compute.
\end{abstract}

\maketitle

Analyzing the correlation functions of emitted signals is a standard tool across different physical disciplines to characterize the properties of atomic, molecular, and solid-state systems~\cite{mandel1995optical,kiilerich2014estimation,kiilerich2016bayesian,yang2023efficient,radaelli2026parameter,mattes2025designing}. The characteristic timescales of correlation functions
are particularly relevant for classical~\cite{cardy1996scaling} and quantum~\cite{sachdev2011quantum} systems undergoing phase transitions, as dynamic correlations of order parameters exhibit universal scalings near the critical point. However, fundamental and practical constraints limit our capacity to infer correlation functions, both experimentally and theoretically.

Experimentally, the measurement of, e.g., the second-order intensity autocorrelation function of light $g^{(2)}(\tau)$ (\eqref{eq:g2}) is limited by the resolution time of photodetectors. Time-resolved measurement of $g^{(2)}(\tau)$ is routinely performed when its width is in the nanosecond scale, to characterize emitters such as quantum dots~\cite{arakawa2020progress}, single molecules~\cite{toninelli2021single}, and diamond color centers~\cite{doherty2013nitrogen}. Substantial efforts~\cite{salem2013application,hadfield2023single,horoshko2025time} are dedicated to resolving correlations below the 50~ps resolution of photodetectors. 
Theoretically, the analysis of quantum many-body models is often an intractable task: even in a steady state and when the quantum regression theorem can be applied~\cite{breuer2002theory}, extracting correlation times requires the analytical solution or numerical evaluation of the system's dynamics. Alternative approaches tend to lack universality and often rely on specialized analytical methods. For example, the fluctuation-dissipation theorem~\cite{kubo1966fluctuation} relates symmetrized correlation functions to the response function, which can be evaluated using the Kubo formula~\cite{albert2016geometry}. However, this approach assumes thermalization, a condition frequently violated in, e.g., driven-dissipative quantum systems~\cite{Minganti18}. The Keldysh formalism~\cite{kamenev2023field,sieberer2016keldysh} may be used to compute correlation functions, such as $g^{(1)}(\tau)$ or $g^{(2)}(\tau)$, and their timescales~\cite{zhang2021driven}. Obtaining these results involves a highly nontrivial analytical treatment grounded in quantum field theory and relies on approximations that are generally not controllable.

\begin{figure}[t!]
  \includegraphics[width=0.35
\textwidth]{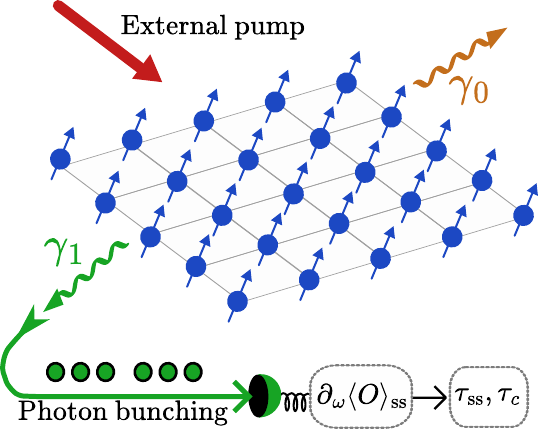}
\centering
\caption{We consider a general open quantum system coupled to a Markovian environment, possibly subject to an external pump. The total emission rate is ${\gamma = \gamma_0 + \gamma_1}$, where $\gamma_0$ corresponds to dissipation into unmonitored modes, while $\gamma_1$ represents the fraction of output photons that are collected and measured. We focus on the steady-state expectation value $\langle O \rangle_{\rm ss}$ of an observable $O$, which depends on the system parameter $\omega$. Using recent theoretical results for quantum parameter estimation, we demonstrate that the sensitivity $\partial_\omega\langle O \rangle_{\rm ss}$ to small variations in $\omega$ provides a lower bound to the relaxation time ($\tau_{\rm ss}$) and, if $O$ is proportional to the number of emitted photons, also to the second-order correlation time ($\tau_{\rm c}$). Our method enables the estimation of the time correlations based solely on the single-time expectation values of the system’s steady state, as a function of $\omega$.}
\label{fig:1}
\end{figure}

In this work, we establish a universal relationship between a system's steady-state sensitivity to small changes of
a system parameter and the second-order correlation function $g^{(2)}(\tau)$, using information-theoretic bounds from quantum parameter estimation~\cite{fujiwara2008fibre,escher2011general,demkowicz2012elusive,knysh2014true,demkowicz2014using,zhou2021asymptotic,kurdzialek2022using,zhou2018achieving,das2024universal,demkowicz2017adaptive,wan2022bounds,gorecki2024interplay}. In particular, our result provides rigorous lower bounds on the correlation and relaxation times. The main idea is that, in continuous steady-state measurements, the sensitivity of an observable with respect to the estimated parameter can be quantified via the signal-to-noise ratio (SNR), which is bounded from above by quantum Fisher information (QFI). If the measurements were uncorrelated in time, the total SNR would accumulate additively, resulting in a precision consistent with the statistics of independent samples. Instead, if the precision seems to exceed the fundamental bounds for quantum parameter estimation obtained under this assumption, we must conclude that the measurement outcomes are temporally correlated. These correlations are captured by the two-time function $g^{(2)}(\tau)$, and we show that such metrological tools can be used to lower bound its characteristic time.

Our bound has both theoretical and experimental implications: it serves both as a conceptual framework for the theoretical analysis of many-body quantum systems and as a practical method for experimentally probing the correlation times of ultrafast dynamics that cannot be resolved by available detectors. Indeed, it enables the estimation of the correlation time based solely on the single-time expectation values of the system’s steady state, without requiring resolution of the dynamics. This method can be readily extended to the autocorrelation of arbitrary observables, and it particularly fits systems showing strong bunching and large susceptibility, such as critical systems~\cite{zanardi2006ground, rams2018limits,Garbe2020}.

We benchmark our method on two complementary examples. First, we consider a parametrically driven resonator, where the bound can be saturated (up to the leading order). This validates the approach and emphasizes its relevance for real-world scenarios, as confirmed by recent experiments~\cite{beaulieu2025observation,beaulieu2025criticality}. Second, we study the infinite-range dissipative transverse-field Ising model, for which a direct solution of the equations of motion is not available. In this case, our method provides new insight into the system’s correlation behavior that would otherwise be hard to access.

{\it Setting and fundamental bounds---} 
As sketched in \autoref{fig:1}, we consider a system governed by the Markovian master equation
\begin{equation}
\begin{split}
\label{eq:motion}
    \frac{d\rho}{dt}=-i[H,\rho]+\gamma\mathcal  D[\rho],\,\, \t{with}\,\, H=\omega G_0+G_1,\,G_0\geq 0,
\end{split}
\end{equation}
where $\omega G_0$ is the free Hamiltonian of the system and $G_1$ is a driving part. The $G_0$ has integer spectrum and is positive~\footnote{Note that the $G_0\geq 0$ condition may always be trivially satisfied by adding a term proportional to $\openone$ without affecting the dynamics}. The Lindbladian is ${\mathcal D[\rho]=\sum_k D_k\rho D_k^\dagger-\frac{1}{2}\{D^\dagger_kD_k,\rho\}}$, where $D_k$ are the lowering operators, mapping between eigenstates of $G_0$ corresponding to neighboring eigenvalues, such that ${G_0= \sum_k D_k^\dagger D_k}$. There is no assumption about $G_1$. This situation arises in many open-driven systems \cite{chen2023quantum,roberts2023exact,alushi2024optimality} (see also \textit{Examples} section). Note that the steady state of such a system $\rho_{\ss}$ is mixed but is not thermal, and the eigenbasis of $\rho_{\ss}$ does not correspond to the eigenbasis of the full Hamiltonian.
Quantum jumps correspond to photon emission into the environment~\cite{kiilerich2014estimation}. The mean emission rate is ${\gamma\braket{G_0}}$, and we denote the emitted mode by $b$.

In what follows, we relate characteristic dynamical times to the steady-state sensitivity to small changes of $\omega$ using quantum-metrological arguments. We consider an estimation protocol in which the system is initially prepared in a state independent of the actual value of $\omega$, and information about $\omega$ is encoded in the system state (and in the emitted light) during the subsequent evolution. Let $O$ denote a measurable scalar quantity used to estimate $\omega$—for instance, an observable measured on the system at time $t$, or a time-integrated observable of the emitted field. The usefulness of $O$ is quantified by its signal-to-noise ratio, $\t{SNR}=\frac{|\partial_\omega \braket{O}|^2}{\Delta^2O}$. The maximal SNR is bounded by the quantum Fisher information, $\t{SNR}\leq \t{QFI}$, both defined with respect to $\omega$.

Let us decompose the total photon-loss (emission) rate as ${\gamma = \gamma_0 + \gamma_1}$, where $\gamma_1\in[0,\gamma]$ denotes the rate of the detected (accessible) output and $\gamma_0$ the rate of light lost to the environment (inaccessible). The maximal QFI achievable over a time $t$ is fundamentally bounded by~\cite{demkowicz2017adaptive,wan2022bounds,gorecki2024interplay} (see Supplemental Material Section~\ref{app:bound}):

\begin{equation}
\label{eq:bound1}
\t{QFI}\leq \frac{4}{\gamma_0}\int_0^t\braket{G_0}_{t'} dt',
\end{equation}
where $\braket{G_0}_{t'}=\tr(\rho(t')G_0)$.
Notably, this bound depends only on the rate of photons lost to the environment $\gamma_0$, not on the total emission rate $\gamma$.

Trivially, for the best precision in estimating $\omega$, one should measure all emitted photons, i.e., setting $\gamma_1=\gamma$ and $\gamma_0=0$. However, in that case, the bound \eqref{eq:bound1} diverges and becomes non-informative~\footnote{That is because, in the absence of losses, Heisenberg scaling of the precision with time is in principle achievable, and therefore a different version of the bound should be used~\cite{demkowicz2017adaptive,wan2022bounds,gorecki2024interplay}. Nevertheless, such a bound would not be useful for our considerations.}. Since our aim is not to find an optimal strategy for estimating $\omega$, but rather to establish a relation between metrological bounds and characteristic dynamical times, ${\gamma_1}$ is only a theoretical tool, and it is treated as a free parameter.

{\it Main results---} In the following, we derive bounds on the relaxation time $\tau_{\ss}$, which is the time required to reach the steady state, and on the second-order correlation time $\tau_{\rm c}$ (\eqref{eq:cortime}).
First, consider the case in which, after reaching a steady state, one performs a direct (destructive) measurement in the eigenbasis of an observable $O$ and attempts to estimate $\omega$ from its average value. We introduce ${\braket{G_0}_{\max}=\max_{t'\in[0,\tau_{\ss}]}\braket{G_0}_{t'}}$, so, from \eqref{eq:bound1}, $\t{QFI}\leq 4\tau_\ss \braket{G_0}_{\max}/\gamma_0$. We set $\gamma_0=\gamma,\gamma_1=0$ (since here no information is extracted from emitted photons), use the inequality $\t{SNR}\leq \t{QFI}$, and rearrange the terms to get:
\begin{equation}
\label{eq:boundtss}
\tau_{\ss}\geq \frac{\gamma}{4\braket{G_0}_{\max}}\frac{|\partial_\omega \braket{O}_{\rm ss}|^2}{\Delta_\ss^2 O}.
\end{equation}
Note that, for finite-dimensional systems, $\braket{G_0}_{\max}$ is bounded by the operator norm $\braket{G_0}_{\max}\leq \|G_0\|$~\footnote{The operator norm of a matrix is its largest singular value.}.

Next, consider the case in which, after obtaining the steady state, one wants to estimate $\omega$ from observing emitted photons. If we consider a very long time ${t\gg \tau_{\ss}}$, the integral in \eqref{eq:bound1} approximates $t\braket{G_0}_\ss$, so
\begin{equation}
\label{eq:bound}
\t{QFI}/t\leq \frac{4\braket{G_0}_{\ss}}{\gamma_0}.
\end{equation}
To properly describe the time correlation, we switch now to the Heisenberg picture, choosing as $t=0$ an arbitrary time after attaining the steady state, i.e. $\rho(0)=\rho_\ss$. For any observables $A,B$, we have $\braket{A(t)B(t')}_{\ss}=\tr(A(t)B(t')\rho_\ss)$, where $A(t),B(t')$ are obtained vie the action of the adjoint Liouvillian super-operator~\cite{roberts2021hidden} associated with \eqref{eq:motion}. Single-time quantities like $\braket{A(t)}_{\ss}$ do not depend on the actual value of $t$ and are therefore denoted simply as $\braket{A}_{\ss}$.
One aims to estimate $\omega$ from the average number of photons counted in the time interval $t$.
For this particular quantity, the
SNR formula per unit time is given by (see Supplemental Material Section~\ref{app:correlation})
\begin{equation}\label{eq:main}
    \t{SNR}/t=\frac{|\partial_{\omega} \gamma_1\braket{G_0}_\ss|^2}{\gamma_1\braket{G_0}_\ss+\gamma_1^2\braket{G_0}_\ss^2\int_{-\infty}^{+\infty}(g^{(2)}(\tau)-1)d\tau},
\end{equation}
where $g^{(2)}(\tau)$ is the normalized, normally ordered second-order correlation function in the steady state:
\begin{equation}
\label{eq:g2}
g^{(2)}(\tau)=\frac{\braket{\bb^\dagger(0)\bb^\dagger(\tau)\bb(\tau)\bb(0)}_\ss}{\braket{\bb^\dagger\bb}_\ss^2},
\end{equation}
where $b(t)$ is the output mode~\cite{navarrete2022introduction,walls2007quantum}.
We define the second-order correlation time as 
\begin{equation}
\label{eq:cortime}
{\tau_{\rm c}\coloneqq\int_{-\infty}^{+\infty} [g^{(2)}(\tau)-1]d\tau}.
\end{equation}
This definition generalizes the standard integral correlation time of classical stationary stochastic processes \cite{Gardiner2009,VanKampen2007} to the quantum domain, measuring the typical timescale over which photon-number fluctuations are correlated.

Using inequality $\t{SNR}/t\leq \t{QFI}/t$, the upper bound for QFI \eqref{eq:bound} and $\gamma_0=\gamma-\gamma_1$ we therefore obtain:
\begin{equation}
    \frac{|\partial_\omega\braket{G_0}_{\rm ss}|^2}{\frac{\braket{G_0}_{\rm ss}}{\gamma_1}+\braket{G_0}_{\rm ss}^2\tau_{\rm c}}\leq \frac{4\braket{G_0}_{\rm ss}}{\gamma-\gamma_1}.
\end{equation}
Rearranging the terms yields a family of bounds for $\tau_{\rm c}$:
\begin{equation}
\tau_{\rm c}\geq 
 \frac{1}{\braket{G_0}_\ss}\left[\frac{(\gamma-\gamma_1)|\partial_\omega\braket{G_0}_\ss|^2}{4\braket{G_0}_{\rm ss}^2}-\frac{1}{\gamma_1}\right].
\end{equation}
The correlation time $\tau_{\rm c}$ depends only on the system dynamics \eqref{eq:motion} and not on $\gamma_1$, therefore the tightest version of the bound may be obtained by maximizing right-hand-side over ${\gamma_1\in[0,\gamma]}$, what results in ${\gamma_1=\min\{2\langle G_0 \rangle_{\ss}/|\partial_\omega \langle G_0 \rangle_{\ss}|,\gamma\}}$. Assuming  $2\langle G_0 \rangle_{\ss}/|\partial_\omega \langle G_0 \rangle_{\ss}|\leq \gamma$ (which is typically satisfied at, e.g, the critical point of a quantum system undergoing a phase transition), this yields the following bound expressed solely in terms of system parameters:
\begin{equation}
\label{eq:corrbound}
 \tau_{\rm c}\geq 
 \frac{1}{\braket{G_0}_\ss}\left[\frac{\gamma|\partial_{\omega} \braket{G_0}_\ss|^2}{4\braket{G_0}_\ss^2}-\frac{|\partial_\omega\braket{G_0}_\ss|}{\braket{G_0}_\ss}\right].
\end{equation}

Even though a suitable choice of $G_1$ in \eqref{eq:motion} may induce critical behavior, it enters the bound only indirectly through its effect on $\braket{G_0}_\ss$. Near criticality, the first term on the right-hand side of \eqref{eq:corrbound} typically dominates and becomes arbitrarily large, signaling strong bunching in the output light (recall that $\tau_{\rm c}$ reflects both the bunching strength and the correlation decay time).

Although derived within a metrological context, the bounds remain valid in any framework. In particular, we can consider a situation where $\omega$ is a known, controllable parameter. The bound can then be evaluated directly from experimental data, without explicitly measuring time correlations. Notice that the decomposition $\gamma=\gamma_0+\gamma_1$ is a purely theoretical construct used to derive the bound in \eqref{eq:corrbound}, which depends only on $\gamma$. Therefore, no experimental tuning of $\gamma$ is required, but only the knowledge of its value. Both $\braket{G_0}_\ss$ and $|\partial_\omega \braket{G_0}_\ss|$ can be easily estimated by varying $\omega$, and therefore also the bound in \eqref{eq:corrbound}.

The bound is particularly advantageous when $g^{(2)}(\tau)$ cannot be measured directly, e.g., due to detector dead time exceeding the relaxation time, or in the weak-coupling and limited-acquisition regime (see Supplemental Material Section~\ref{app:weakcoupling}).

\begin{figure}[t!]
  \includegraphics[width=0.46
\textwidth]{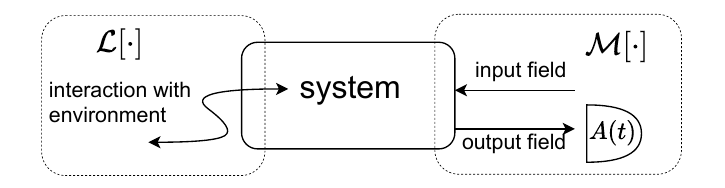}
\centering
\caption{Schematic representation of an open quantum system driven by an input field, with continuous measurement performed on the output field, described by \eqref{general_eq}. 
}
\label{fig:general}
\end{figure}

{\it Arbitrary parameter and observable of the output field---}
The above discussion can be generalized to an arbitrary system Hamiltonian parameter $\nu$ and arbitrary observable $A$.
Consider the master equation
\begin{equation}\label{general_eq}
    \frac{d\rho}{dt}=-i[H(\nu),\rho]+\mathcal  L[\rho]+\mathcal M[\rho],
\end{equation}
where $H(\nu)$ is the full Hamiltonian of the system, ${\mathcal L[\rho]:=\sum_k L_k\rho L_k^\dagger-\frac{1}{2}\{L_k^\dagger L_k,\rho \}}$ describes the interaction with an environment and $\mathcal M[\rho]$ accounts for all effects related to the interaction with a controlled field (see \autoref{fig:general}).

If the operator $\partial_\nu H$ can be expressed as a linear combination of identity, Lindblad operators, and their products ---i.e., ${\partial_\nu H\in\t{span}\{\openone,L_i,L_i^\dagger,L_i^\dagger L_j\}}$--- then the SNR per unit time can be bounded by \cite{demkowicz2017adaptive,wan2022bounds,gorecki2024interplay}:
\begin{equation}
    \t{SNR}/t\leq C[\partial_\nu H,\{L_i\}_i],
\end{equation}
where $C$ is a scalar constant that depends only on $\partial_\nu H$ and $\{L_i\}_i$.
Then, for the time-integrated observable $O=\int_0^tA(t')dt'$ measured on the output field in the steady state, for large $t$, the SNR per unit time 
$(|\partial_\nu O|^2/\Delta^2O)/t$ may be expanded as:
\begin{equation}
  \frac{|\partial_\nu \braket{A}_{\ss}|^2}{\int_{-\infty}^{+\infty} \left[\braket{A(0)A(t)}_{\ss}-\braket{A}_{\ss}^2\right]dt}\leq  C[\partial_\nu H,\{L_i\}_i],
  \label{qui}
\end{equation}
so if $\partial_\nu \braket{A}_{\ss}$ and $C[\partial_\nu H,\{L_i\}_i]$ are known or measurable directly, it constitutes the lower bound for autocorrelation of $A(t)$. Further expanding \eqref{qui} into an explicit bound on the correlation time requires choosing a specific model; once the model is specified, $C[\partial_\nu H,\{L_i\}_i]$ can be written explicitly.
E.g., the reasoning used to derive \eqref{eq:corrbound} can be seen as a special case of the above with $\nu=\omega$, $A(t)=b_1^\dagger(t)b_1(t)$ (see Supplemental Material Section~\ref{app:correlation}), ${\mathcal L[\cdot]=\gamma_0 \mathcal D[\cdot]}$ and ${\mathcal M[\cdot]=\gamma_1 \mathcal D[\cdot]}$. 
Equations \hyperref[eq:boundtss]{(\ref{eq:boundtss})}, \hyperref[eq:corrbound]{(\ref{eq:corrbound})} and \hyperref[qui]{(\ref{qui})} are the main results of our paper.

{\it Examples---} We illustrate the validity and applicability of our approach through two representative examples. Example (i) is a single-mode degenerate parametric amplifier, described by a squeezing Hamiltonian with losses~\cite{walls2007quantum}. This quantum optical system admits analytical solutions for both the steady state and the full dynamics, allowing a verification of our bounds against exact results. Example (ii) is the dissipative infinite-range Ising model with a transverse field, for which an analytical steady-state solution has been recently derived~\cite{roberts2023exact}; however, its dynamics is not analytically tractable. This case illustrates the effectiveness of our method in exploring regimes that are beyond the reach of conventional analytical techniques.

{\it (i) Lossy cavity with parametrically induced squeezing---}
Consider a single-mode cavity governed by a squeezing Hamiltonian and coupled to the vacuum bath, i.e.:
\begin{equation}
\label{eq:sqham}
 H=\omega a^\dagger a+\frac{\epsilon}{2}(a^{\dagger2}+a^2),\quad \mathcal D[\rho]=a\rho a^\dagger-\frac{1}{2}\{a^\dagger a,\rho\}.
\end{equation}
This Hamiltonian can be effectively realized by driving a nonlinear resonator with a pump at approximately twice the cavity frequency~\footnote{In this implementation, the effective squeezing parameter is proportional to the pump strength times the nonlinearity coefficient, and inversely proportional to the cavity frequency; see Eq. (2) and App. A of Ref.~\cite{frattini2024observation} for details.}. The required nonlinearity can be engineered using suitable Josephson-junction circuits~\cite{yamamoto2008,Zhong_2013,frattini2024observation}. Far from the critical threshold, the static nonlinearity is negligible, yielding the quadratic Hamiltonian \eqref{eq:sqham}. The detuning $\omega$ can be controlled via magnetic tuning of SQUID elements~\cite{Krantz_2013}.

For ${\epsilon<\sqrt{\omega^2+(\gamma/2)^2}}$, the system steady state is a Gaussian state with zero mean and an average number of photons given by ${ N_{\ss}=\epsilon^2/[2(\omega^2+(\gamma/2)^2-\epsilon^2)]}$.
As it has been recently observed in experiment~\cite{beaulieu2025observation,beaulieu2025criticality}, the system exhibits critical behavior when $\epsilon$ approaches the critical value ${\epsilon_c=\sqrt{\omega^2+(\gamma/2)^2}}$, and in that regime we have ${\partial_\omega N_\ss\approx -\frac{4\omega}{\epsilon_c^2}N_\ss^2}$, and ${\Delta^2N_\ss\approx 2N^2_\ss}$. Here ${G_0=a^\dagger a}$ and ${\braket{G_0}_\ss=\braket{a^\dagger a}_\ss=N_\ss}$.
Assuming ${\braket{G_0}_{\max}=\braket{G_0}_\ss}$, which is always true for ${\epsilon>\omega}$ ~\cite{alushi2024optimality}, from \eqref{eq:boundtss} with ${O=G_0}$ we obtain
\begin{equation}
\tau_{\ss}\geq \frac{2\omega^2\gamma}{\epsilon_c^4}N_\ss,
\end{equation}
which is consistent with the analytical solution, as the inverse of the lowest nonzero real part of the Liouvillian eigenvalue also scales linearly with $N_\ss$~\cite{alushi2024optimality}. 

For the correlation function, by substituting the exact formulas for $\braket{G_0}_\ss$ and $\partial_\omega \braket{G_0}_\ss$ into \eqref{eq:corrbound}, we obtain
\begin{equation}
\label{eq:cavitybound}
\tau_{\rm c}\geq \frac{2\omega^2\gamma}{\epsilon^2}\frac{1}{\omega^2+(\gamma/2)^2-\epsilon^2}-\frac{4\omega}{\epsilon^2}\approx 4\frac{\omega^2\gamma}{\epsilon_c^4}N_{\ss},
\end{equation}
so bunching of the emitted photons scales linearly with the average photon number~\footnote{For comparison, the correlation function for thermal light in the cavity is ${g_{\t{thermal}}^{(2)}(\tau)=1+e^{-\gamma \tau}}$~\cite[Eq. (530)]{navarrete2022introduction}, so ${\tau_{\rm c}^{\t{thermal}}=\frac{2}{\gamma}}$, does not depend on the number of thermal photons.}.
Since the model is analytically solvable, one can compute $\tau_{\rm c}$ directly and obtain $\tau_{\rm c}\approx  \frac{4\gamma}{\epsilon_c^2}N_{\ss}$ (see Supplemental Material Section~\ref{app:corcav}). We observe that our bound is tight in the regime ${\omega\gg \gamma}$, whereas it becomes ineffective for ${\omega\ll \gamma}$. In intermediate regimes, the bound provides an accurate estimate of the order of magnitude.

{\it (ii) Dissipative transverse-field Ising model---}
Consider a system of $N$ spin-$1/2$ particles with the Hamiltonian
\begin{equation}
H=\omega\sum_{i=1}^N\sigma^z_i+\sum_{i=1}^Nh_i^x\sigma^x_i+\sum_{i\neq j}J_{ij}\sigma^z_i\sigma^z_j,
\end{equation}
where the terms $h_i^x\sigma^x_i$ represent an external pump, and $J_{ij}\sigma^z_i\sigma^z_j$ accounts for interactions between spins. For now, we make no further assumption of the range of interaction. The dissipative part
\begin{equation}
\mathcal D[\rho]=\gamma\sum_{i=1}^N\left[\sigma_i^-\rho\sigma_i^{-\dagger}-\tfrac{1}{2}\{\sigma_i^{-\dagger}\sigma_i^-,\rho\}\right]
\end{equation}
describes spontaneous emission in the $z$ basis, with each decay corresponding to the emission of a single photon.

To use the bounds, we first define  ${G_0=\sum_{i=1}^N\sigma^z_i+\frac{N}{2}\openone}$ (the shift ensures the lowest eigenvalue of $G_0$ is zero). We define the normalized magnetization operator ${m=G_0/N}$, so that $\braket{m}$ ranges between $[0,1]$. The total flux of emitted photons is ${\gamma N\braket{m}}$. Without any further assumptions, we can bound ${\braket{G_0}_{\max}\leq \|G_0\|=N}$, so that \eqref{eq:boundtss} with ${O=G_0}$ yields for the relaxation time
\begin{equation}
\label{eq:isingtime}
\tau_{\ss}\geq \frac{\gamma}{4N}\frac{|\partial_\omega \braket{m}_{\ss}|^2}{\Delta_\ss^2 m}.
\end{equation}
From \eqref{eq:corrbound}, we obtain for the correlation time
\begin{equation}
\label{eq:isingcorr}
 \tau_{\rm c}\geq \frac{1}{N\braket{m}_\ss}
 \left[\frac{\gamma|\partial_{\omega} \braket{m}_\ss|^2}{4\braket{m}_\ss^2}-\frac{|\partial_{\omega} \braket{m}_\ss|}{\braket{m}_\ss}\right].
\end{equation}
These two formulas are valid for arbitrary values of $h_i^x,J_{ij}$ ---for finite or infinite range of interaction--- with their influence encapsulated in the quantity $\partial_{\omega}\braket{m}_\ss$.

We now focus on the special case of the infinite-range Ising model with uniform couplings $\forall_{ij}J_{ij}=J$ and transverse fields $\forall_i h_i^x=h^x$. While its steady state is known analytically~\cite{roberts2023exact}, the full dynamics remains unsolved. This makes our constraints particularly valuable, as they allow temporal characteristics of the system to be inferred from stationary features alone. The thermodynamic limit is obtained by taking ${N\to\infty}$ keeping ${\bar J\coloneqq NJ}$ constant, so we adopt ${1/\bar J}$ as the natural unit of time. The system exhibits critical behavior and a sharp change of magnetization for a certain value of $\omega$, which we call ${\omega_c\coloneqq\t{argmax}_{\omega} |\partial_\omega \braket{m}_{\ss}|}$.

\begin{figure}[t!]
  \includegraphics[width=0.48
\textwidth]{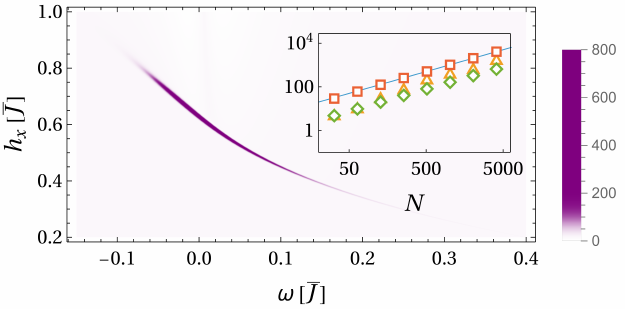}
\centering
\caption{Density plot of the value of the lower bound for output photon correlation $\tau_{\rm c}$ for the infinite-range dissipative transverse-field Ising model
(\eqref{eq:isingcorr}) for fixed loss rate ${\gamma=0.1[\bar J]}$ and number of spins ${N=128}$. The bound takes a significantly large value on the line corresponding to the phase transition.
\textbf{Inset:} The bounds for time correlations $\tau_{\rm c}$ (yellow triangles) and relaxation time $\tau_{\ss}$ (green diamonds), and value of $|\partial_\omega \braket{m}_\ss|$ (red squares) as a function of $N$, calculated for fixed values ${\gamma=0.1[\bar J]}$, ${h^x=0.5[\bar J]}$, at critical point $\omega_c\approx 0.064 [\bar J]$. The blue solid line represents the linear function ${f(N)=N}$ for the reference.
}
\label{fig:ising}
\end{figure}

An exact expression for the magnetization $m$ is known~\cite{roberts2023exact}, allowing both bounds to be written analytically (although the resulting formulas are too lengthy to display). In \autoref{fig:ising}, we show \eqref{eq:isingcorr} over a broad range of $\omega$ and $h^x$ for fixed $\gamma=0.1[\bar J]$ and $N=128$, computed using Eqs. (14) and (S39) of \cite{roberts2023exact}. The high-value ridge in the density plot coincides with the phase transition (cf. Fig. 3(a) of \cite{roberts2023exact}).

In the inset of \autoref{fig:ising}, for fixed ${\gamma=0.1[\bar J]}$ and ${h^x=0.5[\bar J]}$, we analyze the $N$-dependence of the bounds at the critical point ${\omega_c\approx 0.064[\bar J]}$, using the approximation formula (Eq.~(16) of \cite{roberts2023exact}). The steady-state magnetization is ${\braket{m}_{\ss}\approx 0.4}$~\footnote{For the investigated range up to $N=2^{12}$, this value still varies by up to 10\%.}. At criticality, the distribution of $m$ becomes bimodal, so its variance approaches a finite value in the thermodynamic limit, $\Delta_\ss^2m\approx 0.16$. Moreover, $|\partial_{\omega} \braket{m}_{\ss}|\approx N[1/\bar J]$, implying linear scaling of both $\tau_{\ss}$ and $\tau_{\rm c}$ with system size, in line with the behavior observed in the previous example.

This strong bunching contrasts with the single-spin case, where the emitted light is necessarily anti-bunched, since emission projects the system to its ground state~\cite{kiilerich2014estimation}. Applying our bound thus reveals a fundamental dynamical property of an open quantum many-body system.

{\it Conclusions---}
In this work, we established a connection between the correlation time of out-of-equilibrium open quantum systems and their sensitivity to parameter changes, providing lower bounds on the correlation and relaxation times. For a broad class of critical models, we show that the enhanced sensitivity of the mean excitation is intrinsically connected to long relaxation and second-order correlation times. This relation provides a powerful tool for analyzing the second-order correlation function $g^{(2)}(\tau)$ in scenarios where only steady-state properties are accessible, and the full time dynamics are analytically or numerically intractable. In fact, recent results allow one to efficiently estimate steady-state observables numerically, with certified error bounds provided by semidefinite programming methods, even when the steady state itself is not accessible~\cite{PhysRevResearch.7.033237}. 
Our method allows one to lower bound the correlation time using only the average number of photons emitted within a given time interval, measured for two closely spaced values of the system parameter, which makes it particularly useful from an experimental point of view. In this sense, this work extends the theoretical toolbox~\cite{Hauke2016MeasuringMultipartite,Mazza2025EntanglementDetection,Wiesniak2005MagneticSusceptibility,Laurell2025WitnessingEntanglement} for characterizing quantum materials derived in the context of quantum information. The framework we developed can be extended to relate temporal autocorrelations of arbitrary observables to the system’s sensitivity to general Hamiltonian perturbations, opening new avenues for probing quantum criticality in dissipative systems.
Extending these results to non-Markovian environments would be an interesting but nontrivial direction for future research~\cite{kurdzialek2025universal}.

{\it Acknowledgments---} We acknowledge Andrea Smirne, Sauri Bhattacharyya, Valentina Brosco and Jos\'e Lorenzana for insightful discussions. W.G. acknowledges financial support from
the U.S. Department of Energy, Office of Science, National
Quantum Information Science Research Centers, Superconducting Quantum Materials and Systems Center
(SQMS) under the contract No. DE-AC02-07CH11359. S.F. acknowledges financial support from the foundation Compagnia di San Paolo, grant vEIcolo no. 121319, and from
PNRR MUR project PE0000023-NQSTI financed by the European Union – Next Generation EU. R.D. acknowledges financial support from the Academy of Finland, grants  No. 349199, No. 353832, and No. 368477. L.M. acknowledges the PRIN MUR Project 2022RATBS4.

\bibliography{biblio}

@book{VanKampen2007,
  author    = {N. G. van Kampen},
  title     = {Stochastic Processes in Physics and Chemistry},
  edition   = {3},
  series    = {North-Holland Personal Library},
  publisher = {Elsevier},
  year      = {2007},
  doi       = {10.1016/B978-0-444-52965-7.X5000-4}
}

@book{Gardiner2009,
  author    = {C. W. Gardiner},
  title     = {Stochastic Methods: A Handbook for the Natural and Social Sciences},
  edition   = {4},
  series    = {Springer Series in Synergetics},
  publisher = {Springer},
  year      = {2009}
}

@article{roberts2021hidden,
  title={Hidden time-reversal symmetry, quantum detailed balance and exact solutions of driven-dissipative quantum systems},
  author={Roberts, David and Lingenfelter, Andrew and Clerk, AA},
  journal={PRX Quantum},
  volume={2},
  number={2},
  pages={020336},
  year={2021},
  publisher={APS},
doi={10.1103/PRXQuantum.2.020336}
}

@article{Hauke2016MeasuringMultipartite,
  author       = {Philipp Hauke and Markus Heyl and Luca Tagliacozzo and Peter Zoller},
  title        = {Measuring multipartite entanglement through dynamic susceptibilities},
  journal      = {Nat. Phys.},
  volume       = {12},
  pages        = {778--782},
  year         = {2016},
  doi          = {10.1038/nphys3700}
}

@article{Mazza2025EntanglementDetection,
  title = {Entanglement detection in quantum materials with competing orders},
  author = {Mazza, Giacomo and Budroni, Costantino},
  journal = {Phys. Rev. B},
  volume = {111},
  issue = {10},
  pages = {L100302},
  numpages = {7},
  year = {2025},
  month = {Mar},
  publisher = {American Physical Society},
  doi = {10.1103/PhysRevB.111.L100302},
  url = {https://link.aps.org/doi/10.1103/PhysRevB.111.L100302}
}

@article{Wiesniak2005MagneticSusceptibility,
  author       = {M. Wiesniak and V. Vedral and C. Brukner},
  title        = {Magnetic Susceptibility as a Macroscopic Entanglement Witness},
  journal      = {New J. Phys.},
  volume       = {7},
  pages        = {258},
  year         = {2005},
  doi          = {10.1088/1367-2630/7/1/258}
}

@article{Laurell2025WitnessingEntanglement,
  author       = {Pontus Laurell and Allen Scheie and Elbio Dagotto and D. Alan Tennant},
  title        = {Witnessing Entanglement and Quantum Correlations in Condensed Matter: A Review},
  journal      = {Adv. Quantum Technol.},
  volume       = {8},
  pages        = {2400196},
  year         = {2025},
  doi          = {10.1002/qute.202400196}
}

@article{Minganti18,
  title = {Spectral theory of Liouvillians for dissipative phase transitions},
  author = {Minganti, Fabrizio and Biella, Alberto and Bartolo, Nicola and Ciuti, Cristiano},
  journal = {Phys. Rev. A},
  volume = {98},
  issue = {4},
  pages = {042118},
  numpages = {13},
  year = {2018},
  month = {Oct},
  publisher = {American Physical Society},
  doi = {10.1103/PhysRevA.98.042118},
  url = {https://link.aps.org/doi/10.1103/PhysRevA.98.042118}
}

@article{zanardi2006ground,
  title = {Ground state overlap and quantum phase transitions},
  author = {Zanardi, P. and Paunkovi\ifmmode \acute{c}\else \'{c}\fi{}, N.},
  journal = {Phys. Rev. E},
  volume = {74},
  issue = {3},
  pages = {031123},
  numpages = {6},
  year = {2006},
  month = {Sep},
  publisher = {American Physical Society},
  doi = {10.1103/PhysRevE.74.031123},
  url = {https://link.aps.org/doi/10.1103/PhysRevE.74.031123}
}

@article{Garbe2020,
  title = {Critical Quantum Metrology with a Finite-Component Quantum Phase Transition},
  author = {Garbe, L. and Bina, M. and Keller, A. and Paris, M. G. A. and Felicetti, S.},
  journal = {Phys. Rev. Lett.},
  volume = {124},
  issue = {12},
  pages = {120504},
  numpages = {5},
  year = {2020},
  month = {Mar},
  publisher = {American Physical Society},
  doi = {10.1103/PhysRevLett.124.120504},
  url = {https://link.aps.org/doi/10.1103/PhysRevLett.124.120504}
}

@article{kurdzialek2025universal,
  title={Universal bounds for quantum metrology in the presence of correlated noise},
  author={Kurdzia{\l}ek, Stanis{\l}aw and Albarelli, Francesco and Demkowicz-Dobrza{\'n}ski, Rafa{\l}},
  journal={Phys. Rev. Lett.},
  volume={135},
  number={13},
  pages={130801},
  year={2025},
  publisher={APS},
doi={10.1103/jy3v-wkcb}
}

@article{hadfield2023single,
doi={https://doi.org/10.1364/OPTICA.488853},
  title={Single-photon detection for long-range imaging and sensing},
  author={Hadfield, Robert H and Leach, Jonathan and Fleming, Fiona and Paul, Douglas J and Tan, Chee Hing and Ng, Jo Shien and Henderson, Robert K and Buller, Gerald S},
  journal={Optica},
  volume={10},
  number={9},
  pages={1124--1141},
  year={2023},
  publisher={Optica Publishing Group}
}

@article{salem2013application,
doi={https://doi.org/10.1364/AOP.5.000274},
  title={Application of space--time duality to ultrahigh-speed optical signal processing},
  author={Salem, Reza and Foster, Mark A and Gaeta, Alexander L},
  journal={Advances in Optics and Photonics},
  volume={5},
  number={3},
  pages={274--317},
  year={2013},
  publisher={Optical Society of America}
}

@article{arakawa2020progress,
  title={Progress in quantum-dot single photon sources for quantum information technologies: A broad spectrum overview},
  author={Arakawa, Yasuhiko and Holmes, Mark J},
  journal={Appl. Phys. Rev.},
  volume={7},
  number={2},
  pages={021309},
  year={2020},
  publisher={AIP Publishing},
doi={10.1063/5.0010193}
}

@article{toninelli2021single,
  title={Single organic molecules for photonic quantum technologies},
  author={Toninelli, C and Gerhardt, I and Clark, AS and Reserbat-Plantey, A and G{\"o}tzinger, Stephan and Ristanovi{\'c}, Z and Colautti, M and Lombardi, P and Major, KD and Deperasi{\'n}ska, I and others},
  journal={Nat. Mater.},
  volume={20},
  number={12},
  pages={1615--1628},
  year={2021},
  publisher={Nature Publishing Group UK London},
doi={10.1038/s41563-021-00987-4}
}

@article{doherty2013nitrogen,
  title={The nitrogen-vacancy colour centre in diamond},
  author={Doherty, Marcus W and Manson, Neil B and Delaney, Paul and Jelezko, Fedor and Wrachtrup, J{\"o}rg and Hollenberg, Lloyd CL},
  journal={Phys. Rep.},
  volume={528},
  number={1},
  pages={1--45},
  year={2013},
  publisher={Elsevier},
doi={10.1016/j.physrep.2013.02.001}
}

@article{horoshko2025time,
  title = {Time-resolved second-order autocorrelation function of parametric down-conversion},
  author = {Horoshko, Dmitri B. and Srivastava, Shivang and So\ifmmode \acute{s}\else \'{s}\fi{}nicki, Filip and Miko\l{}ajczyk, Micha\l{} and Karpi\ifmmode \acute{n}\else \'{n}\fi{}ski, Micha\l{} and Brecht, Benjamin and Kolobov, Mikhail I.},
  journal = {Phys. Rev. A},
  volume = {112},
  issue = {2},
  pages = {023703},
  numpages = {13},
  year = {2025},
  month = {Aug},
  publisher = {American Physical Society},
  doi = {10.1103/7ckm-tm3r},
  url = {https://link.aps.org/doi/10.1103/7ckm-tm3r}
}

@article{kubo1966fluctuation,
  title={The fluctuation-dissipation theorem},
  author={Kubo, Rep},
  journal={Rep. Prog. Phys.},
  volume={29},
  number={1},
  pages={255},
  year={1966},
  publisher={IOP Publishing},
doi={10.1088/0034-4885/29/1/306}
}

@article{albert2016geometry,
  title={Geometry and response of {L}indbladians},
  author={Albert, Victor V and Bradlyn, Barry and Fraas, Martin and Jiang, Liang},
  journal={Phys. Rev. X},
  volume={6},
  number={4},
  pages={041031},
  year={2016},
  publisher={APS},
doi={10.1103/PhysRevX.6.041031}
}

@book{kamenev2023field,
  title={Field theory of non-equilibrium systems},
  author={Kamenev, Alex},
  year={2023},
  publisher={Cambridge University Press},
doi={10.1017/9781108769266}
}

@article{sieberer2016keldysh,
  title={Keldysh field theory for driven open quantum systems},
  author={Sieberer, Lukas M and Buchhold, Michael and Diehl, Sebastian},
  journal={Rep. Prog. Phys.},
  volume={79},
  number={9},
  pages={096001},
  year={2016},
  publisher={IOP Publishing},
doi={10.1088/0034-4885/79/9/096001}
}

@article{zhang2021driven,
  title={Driven-dissipative phase transition in a {K}err oscillator: From semiclassical {PT} symmetry to quantum fluctuations},
  author={Zhang, Xin HH and Baranger, Harold U},
  journal={Phys. Rev. A},
  volume={103},
  number={3},
  pages={033711},
  year={2021},
  publisher={APS},
doi={10.1103/PhysRevA.103.033711}
}

@article{Krantz_2013,
	doi = {10.1088/1367-2630/15/10/105002},
	url = {https://doi.org/10.1088/1367-2630/15/10/105002},
	year = 2013,
	month = {oct},
	publisher = {{IOP} Publishing},
	volume = {15},
	number = {10},
	pages = {105002},
	author = {Philip Krantz and Yarema Reshitnyk and Waltraut Wustmann and Jonas Bylander and Simon Gustavsseon and William D Oliver and Timothy Duty and Vitaly Shumeiko and Per Delsing},
	title = {Investigation of nonlinear effects in {J}osephson parametric oscillators used in circuit quantum electrodynamics},
	journal = {New J. Phys.}
}

@article{beaulieu2025Criticality,
  title = {Criticality-Enhanced Quantum Sensing with a Parametric Superconducting Resonator},
  author = {Beaulieu, Guillaume and Minganti, Fabrizio and Frasca, Simone and Scigliuzzo, Marco and Felicetti, Simone and Di Candia, Roberto and Scarlino, Pasquale},
  journal = {PRX Quantum},
  volume = {6},
  issue = {2},
  pages = {020301},
  numpages = {16},
  year = {2025},
  month = {Apr},
  publisher = {American Physical Society},
  doi = {10.1103/PRXQuantum.6.020301},
  url = {https://link.aps.org/doi/10.1103/PRXQuantum.6.020301}
}

@article{beaulieu2025observation,
  title={Observation of first-and second-order dissipative phase transitions in a two-photon driven {K}err resonator},
  author={Beaulieu, Guillaume and Minganti, Fabrizio and Frasca, Simone and Savona, Vincenzo and Felicetti, Simone and Di Candia, Roberto and Scarlino, Pasquale},
  journal={Nat. Comm.},
  volume={16},
  number={1},
  pages={1--10},
  year={2025},
  publisher={Nature Publishing Group},
doi={10.1038/s41467-025-56830-w}
}

@article{mattes2025designing,
  title={Designing open quantum systems for enabling quantum-enhanced sensing through classical measurements},
  author={Mattes, Robert and Cabot, Albert and Carollo, Federico and Lesanovsky, Igor},
  journal={Physical Review Letters},
  volume={135},
  number={23},
  pages={230402},
  year={2025},
  publisher={APS},
  doi={10.1103/5gh9-nmv8}
}

@book{sachdev2011quantum,
  title        = {Quantum Phase Transitions},
  author       = {Sachdev, Subir},
  year         = {2011},
  edition      = {2nd},
  publisher    = {Cambridge University Press},
  address      = {Cambridge},
  isbn         = {9780521514682},
  url          = {https://sachdev.physics.harvard.edu/qptweb/}
}

@book{cardy1996scaling,
  title={Scaling and renormalization in statistical physics},
  author={Cardy, John},
  volume={5},
  year={1996},
  publisher={Cambridge university press}
}

@book{mandel1995optical,
  title={Optical coherence and quantum optics},
  author={Mandel, Leonard and Wolf, Emil},
  year={1995},
  publisher={Cambridge university press}
}

@article{rams2018limits,
  title={At the limits of criticality-based quantum metrology: Apparent super-{H}eisenberg scaling revisited},
  author={Rams, Marek M and Sierant, Piotr and Dutta, Omyoti and Horodecki, Pawe{\l} and Zakrzewski, Jakub},
  journal={Phys. Rev. X},
  volume={8},
  number={2},
  pages={021022},
  year={2018},
  publisher={APS},
doi={10.1103/PhysRevX.8.021022}
}

@article{kiilerich2014estimation,
  title={Estimation of atomic interaction parameters by photon counting},
  author={Kiilerich, Alexander Holm and M{\o}lmer, Klaus},
  journal={Phys. Rev. A},
  volume={89},
  number={5},
  pages={052110},
  year={2014},
  publisher={APS},
doi={10.1103/PhysRevA.89.052110}
}

@article{radaelli2026parameter,
  title={Parameter estimation for quantum jump unraveling},
  author={Radaelli, Marco and Smiga, Joseph A and Landi, Gabriel T and Binder, Felix C},
  journal={Quantum},
  volume={10},
  pages={1993},
  year={2026},
  publisher={Verein zur F{\"o}rderung des Open Access Publizierens in den Quantenwissenschaften},
  doi={10.22331/q-2026-02-02-1993}
}

@article{yang2023efficient,
  title={Efficient information retrieval for sensing via continuous measurement},
  author={Yang, Dayou and Huelga, Susana F and Plenio, Martin B},
  journal={Phys. Rev. X},
  volume={13},
  number={3},
  pages={031012},
  year={2023},
  publisher={APS},
doi={10.1103/PhysRevX.13.031012}
}

@article{kiilerich2016bayesian,
  title={Bayesian parameter estimation by continuous homodyne detection},
  author={Kiilerich, Alexander Holm and M{\o}lmer, Klaus},
  journal={Phys. Rev. A},
  volume={94},
  number={3},
  pages={032103},
  year={2016},
  publisher={APS},
doi={10.1103/PhysRevA.94.032103}
}

@article{frattini2024observation,
  title={Observation of pairwise level degeneracies and the quantum regime of the {A}rrhenius law in a double-well parametric oscillator},
  author={Frattini, Nicholas E and Corti{\~n}as, Rodrigo G and Venkatraman, Jayameenakshi and Xiao, Xu and Su, Qile and Lei, Chan U and Chapman, Benjamin J and Joshi, Vidul R and Girvin, SM and Schoelkopf, Robert J and others},
  journal={Phys. Rev. X},
  volume={14},
  number={3},
  pages={031040},
  year={2024},
  publisher={APS},
doi={10.1103/PhysRevX.14.031040}
}

@article{roberts2023exact,
  title={Exact solution of the infinite-range dissipative transverse-field {I}sing model},
  author={Roberts, David and Clerk, Aashish A},
  journal={Phys. Rev. Lett.},
  volume={131},
  number={19},
  pages={190403},
  year={2023},
  publisher={APS},
doi={10.1103/PhysRevLett.131.190403},
}

@article{gorecki2024interplay,
  title = {Interplay Between Time and Energy in Bosonic Noisy Quantum Metrology},
  author = {G\'orecki, Wojciech and Albarelli, Francesco and Felicetti, Simone and Di Candia, Roberto and Maccone, Lorenzo},
  journal = {PRX Quantum},
  volume = {6},
  issue = {2},
  pages = {020351},
  numpages = {18},
  year = {2025},
  month = {Jun},
  publisher = {American Physical Society},
  doi = {10.1103/PRXQuantum.6.020351},
  url = {https://link.aps.org/doi/10.1103/PRXQuantum.6.020351}
}

@article{navarrete2022introduction,
  title={Introduction to quantum optics},
  author={Navarrete-Benlloch, Carlos},
  journal={arXiv preprint arXiv:2203.13206},
  year={2022},
url={https://doi.org/10.48550/arXiv.2203.13206}
}

@article{chen2023quantum,
  title={Quantum behavior of the {D}uffing oscillator at the dissipative phase transition},
  author={Chen, Q.-M. and Fischer, M. and Nojiri, Y. and Renger, M. and Xie, E. and Partanen, M. and Pogorzalek, S. and Fedorov, K. G. and Marx, A. and Deppe, F. and others},
  journal={Nat. Commun.},
  volume={14},
  number={1},
  pages={2896},
  year={2023},
  publisher={Nature Publishing Group UK London},
  url = {https://www.nature.com/articles/s41467-023-38217-x#citeas},
  doi={10.1038/s41467-023-38217-x}
}

@article{das2024universal,
  title={Universal time scalings of sensitivity in {M}arkovian quantum metrology},
  author={Das, Arpan and G{\'o}recki, Wojciech and Demkowicz-Dobrza{\'n}ski, Rafa{\l}},
  journal={Phys. Rev. A},
  volume={111},
  number={2},
  pages={L020403},
  year={2025},
  publisher={APS},
doi={10.1103/PhysRevA.111.L020403}
}

@article{alushi2024optimality,
  title = {Optimality and {{Noise Resilience}} of {{Critical Quantum Sensing}}},
  author = {Alushi, U. and G{\'o}recki, W. and Felicetti, S. and Di Candia, R.},
  year = {2024},
  month = jul,
  journal = {Phys. Rev. Lett.},
  volume = {133},
  number = {4},
  pages = {040801},
  doi = {10.1103/PhysRevLett.133.040801},
  urldate = {2024-07-26}
}

@article{wan2022bounds,
  title = {Bounds on adaptive quantum metrology under {M}arkovian noise},
  author = {Wan, Kianna and Lasenby, Robert},
  journal = {Phys. Rev. Res.},
  volume = {4},
  issue = {3},
  pages = {033092},
  numpages = {18},
  year = {2022},
  month = {Aug},
  publisher = {American Physical Society},
  doi = {10.1103/PhysRevResearch.4.033092},
  url = {https://link.aps.org/doi/10.1103/PhysRevResearch.4.033092}
}

@article{kurdzialek2022using,
  title = {Using Adaptiveness and Causal Superpositions Against Noise in Quantum Metrology},
  author = {Kurdzia\l{}ek, Stanis\l{}aw and G\'orecki, Wojciech and Albarelli, Francesco and Demkowicz-Dobrza\ifmmode \acute{n}\else \'{n}\fi{}ski, Rafa\l{}},
  journal = {Phys. Rev. Lett.},
  volume = {131},
  issue = {9},
  pages = {090801},
  numpages = {7},
  year = {2023},
  month = {Aug},
  publisher = {American Physical Society},
  doi = {10.1103/PhysRevLett.131.090801},
  url = {https://link.aps.org/doi/10.1103/PhysRevLett.131.090801}
}

@article{zhou2021asymptotic,
  title = {Asymptotic Theory of Quantum Channel Estimation},
  author = {Zhou, Sisi and Jiang, Liang},
  journal = {PRX Quantum},
  volume = {2},
  issue = {1},
  pages = {010343},
  numpages = {25},
  year = {2021},
  month = {Mar},
  publisher = {American Physical Society},
  doi = {10.1103/PRXQuantum.2.010343},
  url = {https://link.aps.org/doi/10.1103/PRXQuantum.2.010343}
}

@Book{breuer2002theory,
  Title                    = {The theory of open quantum systems},
  Author                   = {Breuer, Heinz-Peter and Petruccione, Francesco},
  Publisher                = {Oxford University Press on Demand},
  Year                     = {2002}
}

@Article{demkowicz2012elusive,
  Title                    = {The elusive {H}eisenberg limit in quantum-enhanced metrology},
  Author                   = {Demkowicz-Dobrza{\'n}ski, Rafa{\l} and Ko{\l}ody{\'n}ski, Jan and Gu{\c{t}}{\u{a}}, M{\u{a}}d{\u{a}}lin},
  Journal                  = {Nat. Commun.},
  Year                     = {2012},
  Pages                    = {1063},
  Volume                   = {3},
  Publisher                = {Nature Publishing Group},
  url = {https://www.nature.com/articles/ncomms2067}
}

@Article{demkowicz2014using,
  Title                    = {Using entanglement against noise in quantum metrology},
  Author                   = {Demkowicz-Dobrza{\'n}ski, Rafal and Maccone, Lorenzo},
  Journal                  = {Phys. Rev. Lett.},
  Year                     = {2014},
  Number                   = {25},
  Pages                    = {250801},
  Volume                   = {113},
  Publisher                = {APS},
  url = {https://journals.aps.org/prl/pdf/10.1103/PhysRevLett.113.250801}
}

@Article{demkowicz2017adaptive,
  Title                    = {Adaptive Quantum Metrology under General {M}arkovian Noise},
  Author                   = {Demkowicz-Dobrza\ifmmode \acute{n}\else \'{n}\fi{}ski, Rafa\l{} and Czajkowski, Jan and Sekatski, Pavel},
  Journal                  = {Phys. Rev. X},
  Year                     = {2017},
  Pages                    = {041009},
  Volume                   = {7},
  DOI                      = {10.1103/PhysRevX.7.041009},
  Issue                    = {4},
  Numpages                 = {15},
  Publisher                = {American Physical Society},
  url = {https://journals.aps.org/prx/pdf/10.1103/PhysRevX.7.041009}
}

@Article{escher2011general,
  Title                    = {General framework for estimating the ultimate precision limit in noisy quantum-enhanced metrology},
  Author                   = {Escher, BM and de Matos Filho, RL and Davidovich, L},
  Journal                  = {Nat. Phys.},
  Year                     = {2011},
  Number                   = {5},
  Pages                    = {406--411},
  Volume                   = {7},
  Publisher                = {Nature Research},
  url = {https://www.nature.com/articles/nphys1958}
}

@Article{fujiwara2008fibre,
  Title                    = {A fibre bundle over manifolds of quantum channels and its application to quantum statistics},
  Author                   = {Fujiwara, Akio and Imai, Hiroshi},
  Journal                  = {J. Phys. A: Math. Theor.},
  Year                     = {2008},
  Number                   = {25},
  Pages                    = {255304},
  Volume                   = {41},
  Publisher                = {IOP Publishing},
  url = {http://iopscience.iop.org/article/10.1088/1751-8113/41/25/255304/pdf}
}

@Article{knysh2014true,
  Title                    = {True limits to precision via unique quantum probe},
  Author                   = {Knysh, Sergey I and Chen, Edward H and Durkin, Gabriel A},
  Journal                  = {arXiv:1402.0495},
  Year                     = {2014},
  url = {https://doi.org/10.48550/arXiv.1402.0495}
}

@Book{walls2007quantum,
  Title                    = {Quantum optics},
  Author                   = {Walls, Daniel F and Milburn, Gerard J},
  Publisher                = {Springer Science \& Business Media},
  Year                     = {2007}
}

@Article{zhou2018achieving,
  author    = {Zhou, Sisi and Zhang, Mengzhen and Preskill, John and Jiang, Liang},
  title     = {Achieving the {H}eisenberg limit in quantum metrology using quantum error correction},
  journal   = {Nat. Commun.},
  year      = {2018},
  volume    = {9},
  number    = {1},
  pages     = {78},
  publisher = {Nature Publishing Group},
  url       = {https://www.nature.com/articles/s41467-017-02510-3},
}

@article{Zhong_2013,
doi = {10.1088/1367-2630/15/12/125013},
url = {https://dx.doi.org/10.1088/1367-2630/15/12/125013},
year = {2013},
month = {dec},
publisher = {IOP Publishing},
volume = {15},
number = {12},
pages = {125013},
author = {Zhong, L and Menzel, E P and Di Candia, R and Eder, P and Ihmig, M and Baust, A and Haeberlein, M and Hoffmann, E and Inomata, K and Yamamoto, T and Nakamura, Y and Solano, E and Deppe, F and Marx, A and Gross, R},
title = {Squeezing with a flux-driven {J}osephson parametric amplifier},
journal = {New J. Phys.}
}

@article{yamamoto2008,
    author = {Yamamoto, T. and Inomata, K. and Watanabe, M. and Matsuba, K. and Miyazaki, T. and Oliver, W. D. and Nakamura, Y. and Tsai, J. S.},
    title = {Flux-driven {J}osephson parametric amplifier},
    journal = {Appl. Phys. Lett.},
    volume = {93},
    number = {4},
    pages = {042510},
    year = {2008},
    month = {07},
    issn = {0003-6951},
    doi = {10.1063/1.2964182},
    url = {https://doi.org/10.1063/1.2964182},
}

@article{PhysRevResearch.7.033237,
  title = {Certifying steady-state properties of open quantum systems},
  author = {Mortimer, Luke and Farina, Donato and Di Bello, Grazia and Jansen, David and Leitherer, Andreas and Mujal, Pere and Ac\'{\i}n, Antonio},
  journal = {Phys. Rev. Res.},
  volume = {7},
  issue = {3},
  pages = {033237},
  numpages = {7},
  year = {2025},
  month = {Sep},
  publisher = {American Physical Society},
  doi = {10.1103/hbrt-cn8q},
  url = {https://link.aps.org/doi/10.1103/hbrt-cn8q}
}


\clearpage
\onecolumngrid
\begin{center}
  {\large\bfseries Supplemental Material to: ``Time correlations from steady-state expectation values''}\\[4mm]
  Wojciech Górecki,$^{1}$ Simone Felicetti,$^{2,3}$
  Lorenzo Maccone,$^{1,4}$ and Roberto Di Candia$^{5,2,4}$\\[3mm]
  {\itshape\footnotesize
  $^{1}$INFN Sez.\ Pavia, via Bassi 6, I-27100 Pavia, Italy\\
  $^{2}$Institute for Complex Systems, National Research Council (ISC-CNR),
        Via dei Taurini 19, 00185 Rome, Italy\\
  $^{3}$Physics Department, Sapienza University, P.le A. Moro 2, 00185 Rome, Italy\\
  $^{4}$Dip.\ Fisica, University of Pavia, via Bassi 6, I-27100 Pavia, Italy\\
  $^{5}$Department of Information and Communications Engineering,
        Aalto University, Espoo, 02150 Finland}
\end{center}
\vspace{4mm}
\twocolumngrid

\setcounter{equation}{0}
\setcounter{figure}{0}
\setcounter{table}{0}
\setcounter{section}{0}
\renewcommand{\theequation}{S\arabic{equation}}
\renewcommand{\thefigure}{S\arabic{figure}}
\renewcommand{\thetable}{S\arabic{table}}

\section{Derivation of the bound \eqref{eq:bound1}}
\label{app:bound}

Below, we recall the derivation of the upper bound on the QFI of the joint state of the system and the output light mode after an evolution time $t$, following the method of \cite{wan2022bounds}, leading to \eqref{eq:bound1}.

Let us introduce a more general metrological framework, in which the model discussed in the main text appears as a special case. In addition to the sensing system $\mathcal H_S$, we introduce the meter $\mathcal H_M$. The meter is a quantum system that interacts with the sensing system, acquires information about its state, and is directly accessible to measurement. In this way, the parameter is encoded in the sensing system, while the measurement is performed on the meter. In our model, the meter will play the role of the environmental modes relative to the $\gamma_1$ output channel, corresponding to the fraction of emitted photons that can be measured. We assume that the meter itself is completely noiseless. Since our goal is to derive an upper bound on the QFI, this idealized assumption does not affect the validity of the result: any additional noise acting on the meter can only decrease the total QFI. The joint evolution of the system and the meter is given by the master equation:
\begin{equation}
\label{eq:masterwan}
    \frac{d\varrho}{dt} = -i[H\otimes \openone_M+H_\t{INT}, \varrho] + \gamma_0 \mathcal{D}[\varrho],
\end{equation}
where the parameter $\omega$ is encoded only in $H$. The Lindbladian is ${\mathcal D[\varrho]=\sum_k D_k\varrho D_k^\dagger-\frac{1}{2}\{D^\dagger_kD_k,\varrho\}}$ and $D_k$ acts solely on the system (i.e. $D_k$ is shorthand for $D_k\otimes \openone_M$). Here we introduce $\varrho$ as the joint state of the system and the meter, while $\rho$ appearing in the main text denotes the reduced state of the system.
In particular, in our specific model, the evolution of the reduced density matrix of the sensing system, $\rho=\tr_M(\varrho)$, reproduces~\eqref{eq:motion} with $\gamma=\gamma_0+\gamma_1$. However, the derivation below remains valid for an arbitrary meter $\mathcal H_M$ and system--meter interaction Hamiltonian $H_\mathrm{INT}$.

Let $\varrho':=\frac{d\varrho}{d\omega}$ and $\dot\varrho:=\frac{d\varrho}{dt}$. $\varrho'$ may be expressed using the symmetric logarithmic derivative (SLD) $\Lambda$:
\begin{equation}
\label{eq:sld}
    \varrho'=\frac{1}{2}(\varrho \Lambda+\Lambda\varrho).
\end{equation}
Taking the time derivative gives:
\begin{equation}
\label{eq:sldvar}
    \dot\varrho'=\frac{1}{2}(\dot\varrho \Lambda+\Lambda\dot\varrho+\varrho \dot \Lambda+\dot \Lambda\varrho).
\end{equation}
By definition, the QFI is given by
\begin{equation}
    {\t{QFI}}=\tr(\varrho \Lambda^2).
\end{equation}
Taking the time derivative yields
\begin{equation}
\label{eq:fishder}
    \dot {\t{QFI}}=\tr(\dot\varrho \Lambda^2)+\tr(\varrho \dot \Lambda\Lambda)+\tr(\varrho \Lambda\dot \Lambda).
\end{equation}
Applying $\tr(\cdot \Lambda)$ to \eqref{eq:sldvar}, rearranging the terms and substituting the result into \eqref{eq:fishder}, one obtains
\begin{equation}
\label{eq:fishderfin}
    \dot {\t{QFI}}=2\tr(\dot \varrho' \Lambda)-\tr(\dot \varrho \Lambda^2).
\end{equation}
Now we express the above formula for $\dot{\t{QFI}}$ in terms of $\varrho$, $\frac{dH}{d\omega}$, $D_k$, and $\Lambda$.

Taking the derivative of \eqref{eq:masterwan} with respect to $\omega$ gives
\begin{equation}
\label{eq:masterder}
    \dot \varrho'= -i[H'\otimes \openone_M, \varrho]-i[H\otimes \openone_M+H_\t{INT}, \varrho'] + \gamma_0 \mathcal{D}[\varrho'],
\end{equation}
while $\varrho'$ is given by \eqref{eq:sld}. Substituting \eqref{eq:masterwan} for $\dot \varrho$ and \eqref{eq:masterder} for $\dot\varrho'$ into \eqref{eq:fishderfin} and rearranging the terms, we obtain
\begin{equation}
\label{eq:fishderstep}
    \dot {\t{QFI}}=2i \tr(\varrho[H',\Lambda])-\gamma_0\sum_k \tr(\varrho[D_k,\Lambda]^\dagger [D_k,\Lambda]).
\end{equation}
Note that the terms containing $H\otimes \openone_M+H_\t{INT}$ do not contribute to $\dot{\t{QFI}}$, leaving only the term containing $H'$. This is because the unitary evolution given by the $\omega$-independent part of Hamiltonian neither encodes information about the parameter nor destroys the information already accumulated. Its effect is nevertheless reflected in the instantaneous state of the system, $\varrho(t)$. In contrast, dissipation is irreversible and may reduce the quantum Fisher information, which is why the dissipative operators $D_k$ appear explicitly. 

Now consider the special case $H=\omega G_0+G_1$, where $G_0=\sum_k D_k^\dagger D_k$. Then, using identity:
\begin{equation}
2i[D^\dagger_kD_k,\Lambda]=2i D_k^\dagger [D_k,\Lambda]-2i[D_k,\Lambda]^\dagger D_k,
\end{equation}
the equation \eqref{eq:fishderstep} may be rewritten as
\begin{multline}
    \dot {\t{QFI}}=\frac{4}{\gamma_0}\sum_k \tr(\varrho D_k^\dagger D_k)\\
    -\gamma_0\sum_k \tr\!\left(\varrho\left([D_k,\Lambda]+\frac{2i}{\gamma_0} D_k\right)^\dagger
    \left([D_k,\Lambda]+\frac{2i}{\gamma_0} D_k\right)\right).
\end{multline}
Note that the operator
$\left([D_k,\Lambda]+\frac{2i}{\gamma_0}D_k\right)^\dagger
\left([D_k,\Lambda]+\frac{2i}{\gamma_0}D_k\right)$
is positive semidefinite by construction, so every term in the sum is non-negative. Since the sum is multiplied by the negative coefficient $-\gamma_0$, it follows that
\begin{equation}
    \dot {\t{QFI}}\leq \frac{4}{\gamma_0}\sum_k \tr(\varrho D_k^\dagger D_k)=\frac{4}{\gamma_0}\braket{G_0}.
\end{equation}
Integrating over time gives the bound \eqref{eq:bound1}:
\begin{equation}
\t{QFI}\leq \frac{4}{\gamma_0}\int_0^t\braket{G_0}_{t'}\, dt'.
\end{equation}


The above reasoning allows one to apply the bound to analyze situations in which a fraction of the emitted photons is observed ($\gamma_1$), while another fraction is completely lost ($\gamma_0$). The bound depends exclusively on the number of lost photons ($\gamma_0$), as it remains a valid upper bound on the QFI attainable in any protocol performed on such a system.

In the case where information about the parameter is obtained solely through the observation of emitted photons ($\gamma_1$), the performance of this specific protocol will, of course, also depend on $\gamma_1$. Nevertheless, it cannot exceed the bound. Increasing $\gamma_1$ raises the number of collected photons, but at the same time perturbs the system’s steady state, making it less sensitive to variations in $\omega$. Even if one simultaneously increases both $\gamma_1$ and the pump rate $G_1$, so as to maintain the same average excitation number in the system, the emitted photons will carry less information about the parameter per photon. As a result, the optimal metrological strategy requires a compromise value of $\gamma_1$, which provides a significant amount of information while not perturbing the system excessively.

This is the interpretation of the bound in a metrological context. We emphasize, however, that in this work we are not interested in determining the value of $\gamma_1$ that is optimal for estimating $\omega$, but rather in estimating correlations in the emitted light. Although this task draws on metrological concepts, it is fundamentally a different problem.

\section{Derivation of \eqref{eq:main}}
\label{app:correlation}

To formally express the fact that only a fraction $\gamma_1/\gamma$ of all emitted photons is observed, we define the mode of the detected photons as:
\begin{equation}
\label{eq:outputmode}
    b_1(t)=\sqrt{\gamma_1/\gamma}\,b(t)+\sqrt{1-\gamma_1/\gamma}\,v(t)
\end{equation}
where $b(t)$ denotes the total output mode and $v(t)$ is a vacuum mode. Note that the $g^{(2)}$ function defined in \eqref{eq:g2} is identical for the total output mode $b(t)$ and the detected mode $b_1(t)$.

We consider the time-integrated observable $O=\int_0^t b_1^\dagger(t')b_1(t')dt'$. We aim to evaluate the SNR formula
$\t{SNR}:=\frac{|\partial_\omega O|^2}{\Delta^2 O}$.
The variance of $O$ may be written explicitly as
\begin{equation}
\begin{split}
    \Delta^2 O =& \int_0^tdt' \int_0^tdt'' \Big[\braket{b_1^\dagger(t')b_1(t')b_1^\dagger(t'')b_1(t'')}\\&\quad-\braket{b_1^\dagger(t')b_1(t')}\braket{b_1^\dagger(t'')b_1(t'')}\Big]\\
    =&\int_0^t\braket{b_1^\dagger(t')b_1(t')}dt'\\
   & \quad+\int_0^t dt'\int_0^tdt''\braket{b_1^\dagger(t')b_1^\dagger(t'')b_1(t'')b_1(t')}\\
    &\quad-\int_0^t\braket{b_1^\dagger(t')b_1(t')}dt'\int_0^t\braket{b_1^\dagger(t'')b_1(t'')}dt''.
\end{split}
\end{equation}
For the class of systems under consideration, we have
\begin{equation}
    \forall_t \braket{b_1^\dagger(t)b_1(t)}=\gamma_1\braket{G_0(t)}.
\end{equation}

Assuming the system is in a steady state, the variance simplifies to
\begin{align}
        \Delta^2& O =\,\gamma_1 t  \braket{G_0}_{\ss}+\nonumber\\ \quad &\quad\gamma_1^2\braket{G_0}_{\ss}^2\int_0^t\int_0^t\left[g^{(2)}(|t'-t''|)-1\right]dt'dt'', 
\end{align}
where ${g^{(2)}(|t'-t''|)}$ denotes the normalized, normally ordered, second-order time correlation function.

\begin{figure}[t!]
  \includegraphics[width=0.4
\textwidth]{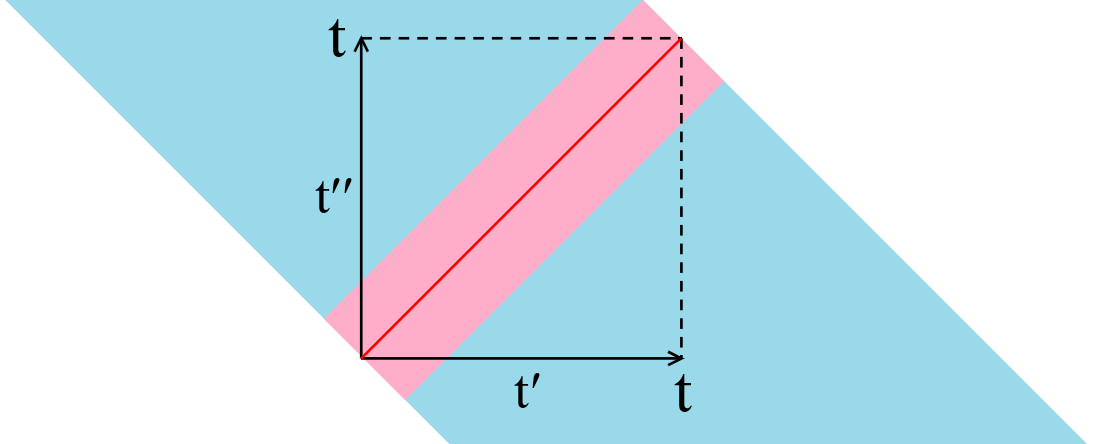}
\centering
\caption{Since the function ${g(|t'-t''|)}$ takes significantly large values only for $t'$ close to $t''$ (pink area), for large $t$ the integral over the square $t\times t$ in \eqref{eq:intappr} can be well approximated by an infinite integral over the blue stripe.}
\label{fig:integral}
\end{figure}

The function ${g(|t'-t''|)-1}$ decays exponentially to zero for large ${|t'-t''|}$, with a decay rate at least equal to the smallest real part of the Liouvillian eigenvalue, $\lambda_{\min}$. Therefore, introducing new variables ${\mathcal{T}=\frac{t+t'}{2}},{\tau={t-t'}}$, the double integral can be approximated as: 
\begin{multline}
\label{eq:intappr}
\int_0^t\int_0^t\left[g^{(2)}(|t'-t''|)-1\right]dt'dt''= \\ 
\\ 
\int\limits_{0}^{t/2} d\mathcal T \int\limits_{-\mathcal T}^{+\mathcal T} \left[g^{(2)}(\tau)-1\right]d \tau+
\int\limits_{t/2}^t d\mathcal T \int\limits_{-(t-\mathcal T)}^{+(t-\mathcal T)} \left[g^{(2)}(\tau)-1\right]d \tau
\\
\overset{t\gg1/\lambda_{\min}}{\approx}
\underbrace{\int_{0}^t d\mathcal T}_{=t} \int_{-\infty}^{+\infty} \left[g^{(2)}(\tau)-1\right]d \tau,
\end{multline}
see \autoref{fig:integral} for a geometrical intuition. Substituting this into the expression for the SNR per unit time yields \eqref{eq:main}.

\section{Weak coupling and limited acquisition time case}
\label{app:weakcoupling}
The bound \eqref{eq:corrbound} is also useful in the case of a perfect detector (zero dead time), in the regime where the average number of photons detected within a time window of length $\sim\tau_{\rm c}$ is much smaller than one --- $\braket{G_0}_\ss\gamma_1\tau_{\rm c} \ll 1$. When estimating $\tau_{\rm c}$, the only informative events are those in which at least two photons are detected within such a time window, which occur with probability ${\approx (\braket{G_0}_\ss\gamma_1\tau_{\rm c})^2}$. One needs many such events to estimate $\tau_{\rm c}$ faithfully, which requires the total acquisition time to satisfy ${t \gg 1/(\braket{G_0}_\ss^2\gamma_1^2\tau_{\rm c})}$.
In contrast, the right-hand side of the bound \eqref{eq:corrbound} depends only on the average photon emission rate. To estimate it, it is sufficient to have ${t \gg \max\{1/(\braket{G_0}_\ss\gamma_1),\tau_{\rm c}\}}$. Hence, for very weak coupling to the detector, the bound can remain informative even when the total acquisition time is insufficient for a direct time-correlation measurement.

\section{Output field for the cavity}
\label{app:cavity}

In this section, we show that for a leaky cavity, the $g^{(2)}(\tau)$ function for the output mode is exactly the same as for the cavity mode. These two are related by~\cite{navarrete2022introduction}:
\begin{equation}
    b(t)=\sqrt{\gamma}a(t)-b_{\t{in}}(t),
\end{equation}
where $b_{\t{in}}(t)$ and $b(t)$ are flux modes (with units of $[1/s]$), while $a(t)$ is the dimensionless, time-dependent cavity mode.

Both $b_{\t{in}}(t)$ and $b(t)$ satisfy standard commutation relation:
\begin{equation}
\begin{split}
    &[b(t),b^\dagger(t')]=\delta(t-t'),
    \quad
    [b(t),b(t')]=0,\\
    &[b_{\t{in}}(t),b^\dagger_{\t{in}}(t')]=\delta(t-t'),
    \quad
    [b_{\t{in}}(t),b_{\t{in}}(t')]=0.
    \end{split}
\end{equation}
The relationship between cavity mode $a(t)$ and input/output modes also satisfies the causality condition:
\begin{equation}
    \forall_{\tau>0}\quad [a^{\dagger}(t-\tau),b_{\t{in}}(t)]=0,\quad 
    [a^{\dagger}(t+\tau),b(t)]=0
\end{equation}
In our case, $b_{\t{in}}(t)$ is assumed to be in the vacuum state. From above, it follows that:
\begin{multline}
\label{eq:inout}
\forall_{\tau>0}
    \braket{b^\dagger(t)b^\dagger(t+\tau)b(t+\tau)b(t)}=\\
\gamma^2\braket{a^\dagger(t)a^\dagger(t+\tau)a(t+\tau)a(t)}.
\end{multline}
Note that this identity holds only for ${\tau>0}$. Indeed, ${a(t+\tau)}$ and ${a(t)}$ do not necessarily commute (see, for instance, \eqref{eq:modedef}). 
Only for ${\tau>0}$, the RHS of the expression has a straightforward physical interpretation--namely, it is proportional to the conditional probability of detecting a photon in the output field at time $t+\tau$, given that one was detected at time $t$.

\section{Time correlation function for lossy single-mode cavity with squeezing Hamiltonian}
\label{app:corcav}
To calculate the second-order correlation function for the model discussed in the Example {\it (i)}, we solve the associated Langevin equation~\cite{walls2007quantum}:
\begin{equation}
\label{eq:langevin}
    \frac{d\boldsymbol{a}(t)}{dt}=A \boldsymbol{a}(t)+\sqrt{\gamma}\boldsymbol{b}_{\t{in}}(t),
\end{equation}
where 
\begin{equation}
\boldsymbol{a}(t)=
\begin{bmatrix}
a(t)\\
a^\dagger(t)
\end{bmatrix},\quad
\boldsymbol{b}_{\t{in}}(t)=
\begin{bmatrix}
b_{\t{in}}(t)\\
b_{\t{in}}^\dagger(t)
\end{bmatrix},
\end{equation}
and $b_{\t{in}}(t)$ is the vacuum input mode satisfying ${[b_{\t{in}}(t),b^\dagger_{\t{in}}(t')]=\delta(t-t')}$). The matrix $A$ is given by
\begin{equation}
   A=\begin{bmatrix}
-i\omega-\gamma/2 & -i\epsilon\\
+i\epsilon & i\omega-\gamma/2. 
\end{bmatrix}
\end{equation}
We postulate a solution of the form
\begin{multline}
\label{eq:modedef}
a(t)=c_1(t)a(0)+c_2(t)a^\dagger(0)+\sqrt{\gamma}\int_0^t c_1(t-\tau)b(\tau)d\tau\\+\sqrt{\gamma}\int_0^t c_2(t-\tau)b^\dagger(\tau)d\tau.
\end{multline}
Substituting this into \eqref{eq:langevin}, we obtain the following system of differential equations:
\begin{equation}
    \begin{split}
        &\dot c_1=-i\epsilon c_2^*-(\gamma/2+i\omega) c_1,\\
        &\dot c_2=-i\epsilon c_1^*-(\gamma/2+i\omega) c_2,
    \end{split}
\end{equation}
with initial conditions $c_1(0)=1$ and $c_2(0)=0$.
The solution is
\begin{multline}
    c_1(x)=\left(\frac{1}{2}-\frac{i\omega}{2\sqrt{\epsilon^2-\omega^2}}\right)e^{-\lambda_-x}+\\\left(\frac{1}{2}
    +\frac{i\omega}{2\sqrt{\epsilon^2-\omega^2}}\right)e^{-\lambda_+x},
\end{multline}
\begin{equation}
    c_2(x)=\frac{-i\epsilon}{2\sqrt{\epsilon^2-\omega^2}}\left(e^{-\lambda_-x}-e^{-\lambda_+x}\right),
\end{equation}
where $\lambda_{\pm}=\gamma/2\pm \sqrt{\epsilon^2-\omega^2}$.

We are interested in computing quantity
\begin{equation}
\lim_{t\to\infty}\braket{a^\dagger(t)a^\dagger(t+\tau)a(t+\tau)a(t)}.
\end{equation}
The contributions involving $c_{1}(t)a(0)$ and  $c_{2}(t)a^\dagger(0)$ vanish in the long-time limit. Since $a(t)$ is a linear combination of both $b(\tau)$ and $b^\dagger(\tau)$, the nonzero contributions come from terms such as
\begin{multline}
\braket{b(t_1)b(t_2)b^\dagger(t_3)b^\dagger(t_4)}=\\
    \delta(t_1-t_4)\delta(t_2-t_3)+\delta(t_1-t_3)\delta(t_2-t_1)
\end{multline}
and
\begin{equation}
\braket{b(t_1)b^\dagger(t_2)b(t_3)b^\dagger(t_4)}=\delta(t_1-t_2)\delta(t_3-t_4)
\end{equation}
appearing under integrals. That leads to:
\begin{equation}
\begin{split}
&\lim_{t\to\infty}\braket{a^\dagger(t)a^\dagger(t+\tau)a(t+\tau)a(t)}=\\ & \gamma^2\cdot\lim_{t\to\infty}\Big[\int_0^t |c_2(t-t')|^2dt'\int_0^{t+\tau} |c_2(t+\tau-t')|^2dt'+\\
    &\int_0^t c^*_2(t-t')c_2(t+\tau-t')dt'\int_0^t c^*_2(t+\tau-t')c_2(t-t')dt'+\\
     &\int_0^t c^*_2(t-t')c^*_1(t+\tau-t')dt'\int_0^t c_1(t+\tau-t')c_2(t-t')dt'
     \Big].
\end{split}
\end{equation}
Since all the integrals involve exponential terms multiplied by constants, they can be evaluated analytically. The final result is
\begin{equation}
\label{eq:g2final}
    \tau_{\rm c}=\frac{2\gamma}{\omega^2+(\gamma/2)^2-\epsilon^2}+\frac{\gamma}{2\epsilon}+\frac{2\omega^2}{\gamma\epsilon^2}.
\end{equation}

To ensure the reader, that the actual value is indeed bigger than the bound for all $\omega,\gamma>0$ and $\epsilon\in(0,\sqrt{\omega^2+(\gamma/2)^2})$, we examine the difference between \eqref{eq:g2final} and \eqref{eq:cavitybound}:
\begin{multline}
   \frac{1}{2\epsilon^2}\left[-3\gamma+8\omega+\frac{4\omega^2}{\gamma}+\frac{4\gamma^3}{\gamma^2+4\omega^2-4\epsilon^2}\right]\geq \\
   \frac{1}{2\epsilon^2}\left[-3\gamma+8\omega+\frac{4\omega^2}{\gamma}+\frac{4\gamma^3}{\gamma^2+4\omega^2}\right]=\\
   \frac{1}{2\epsilon^2}\left[8\omega+\frac{(\gamma^2-4\omega^2)^2}{\gamma(\gamma^2+4\omega^2)}\right]\geq 0,
\end{multline}
where in the first step, we replaced $\epsilon\to0$ in the last term, which may only decrease its value. Finally, near the critical point, the expression simplifies to
\begin{equation}
   \tau_{\rm c}\approx \frac{4\gamma}{\epsilon_c^2}N_{\ss}.
\end{equation}

\end{document}